\documentclass[11pt]{iopart}
\usepackage{iopams}
\usepackage{txfonts}
\usepackage{graphicx}
\usepackage{subfigure}
\usepackage{epsf}
\usepackage{longtable}
\usepackage{lscape}

\newcommand{\aj}{A.J.}
\newcommand{\mnras}{MNRAS}

\newcommand{\msun}{M_{\odot}}

\begin{document}

\article[SCs in BCGs by Adamo et al.]{Star-forming Dwarf Galaxies: Ariadne's Thread in the
  Cosmic Labyrinth}{Tracing the star formation history of three Blue Compact galaxies through the analysis of their star clusters.}

\author{Angela Adamo$^1$, G\"oran \"Ostlin$^1$, Erik Zackrisson$^1$, \& Matthew Hayes$^2$ }

\address{$^1$ Department of Astronomy, Stockholm University, AlbaNova, SE-106 91 Stockholm, Sweden}
\address{$^2$ Observatoire Astronomique de l'UniversitŽ de Genve, 51, ch. des Maillettes,CH-1290 Sauverny, Suisse}

\ead{\mailto{adamo@astro.su.se}}

\begin{abstract}
We present preliminary results from a study of the compact star cluster populations in three local luminous blue compact galaxies: ESO 185-IG 013, ESO 350-IG 038 (a.k.a. Haro\,11), and MRK\,930. These systems show peculiar morphologies and the presence of hundreds of SCs that have been produced by the past, recent, and/or current starburst phases. We use a complete set of HST images ranging from the UV to IR for each galaxy. Deep images in V (WFPC2/f606w) and I (WFPC2/f814w) are used to capture most of the star cluster candidates up to the old ones (fainter) which have had, in the past, less possibility to be detected. The other bands are used in the SED fitting technique for constraining ages and masses. Our goals are to investigate the evolution of these three blue compact galaxies and the star cluster formation impact on their star formation history.
\end{abstract}



\section{Introduction}

Since Searle \& Sargent ({\cite{SS1972}})  for the first time wrote about the discovery of extremely metal-poor galaxies, many studies has been devoted to blue compact galaxies (BCGs). These galaxies appear to be irregular, with total mass and metallicity lower than typical galactic values, blue colors and strong nebular emission lines,  and prominent starburst activities (see {\cite{ostlin01}}, and many others). Thanks to these properties, the BCGs are good candidates for being the building blocks (or at least may have something in comon with the early building blocks) of more massive galaxies, as predicted by the hierarchical galaxy formation paradigm. BCGs were long regarded as candidates for genuinely young galaxies, presently forming in the local universe, but a decisive result in this respect was the detection of red halos (old stellar population) surrounding the starburst region (see {\cite{Kunth00}} for a review). Most BCGs must therefore be old systems experiencing rejuvenation in a new active starburst phase.  

What mechanism trigger the starbursts in BCGs; and moreover -- what supplies the fuel?

\begin{figure}[h]
\begin{center}
\includegraphics[width=12cm]{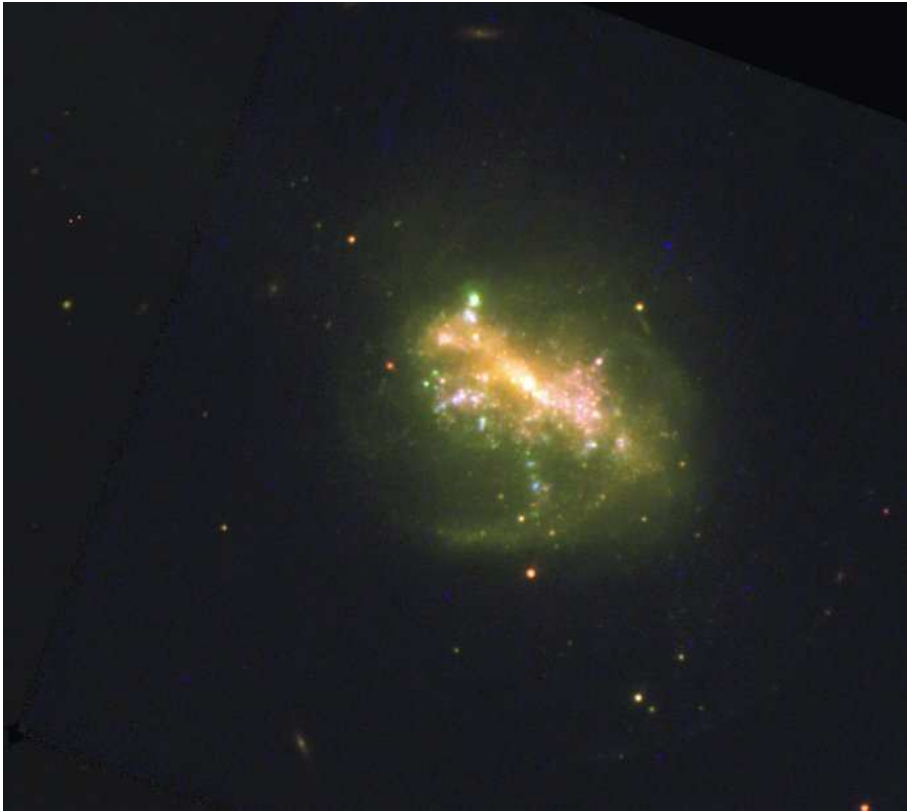}
\caption{Three-color image of the galaxy ESO185 (ESO 185-IG 013). In red WFPC2/f814w filter, in green WFPC2/f606w filter and in blue WFPC2/f336w filter. In \"Ostlin et al. ({\cite{ostlin01}}) this system has been classified as a perturbed merging BCG. The new HST images show clearly the presence of a tidal shell on the south-estern part containing several clusters, and a jet-like structure in the western side (though not so obvious on this intensity scale) . Hundreds of clusters are located in the main body of the galaxy in a kind of barred structure with arm features departing from it.}
\label{eso185}
\end{center}
\end{figure} 
Many models have been proposed: one proposal is that star formation has been reactivated by the encounter of an evolved system and a gas-rich dwarf galaxy ({\cite{lacey91}}). This model can also explain why many BCGs are able to produce a large number of compact star clusters {\cite{ostlin03}}.  
Compact star clusters (SCs) are extremely condensed and bright stellar systems, with a typical mass range of $10^4 - 10^6 \msun$ and small effective radii. In particular, star clusters with masses larger than $10^5 \msun$ are often called super star clusters (SSCs) and may closely resemble the progenitors of  evolved globular clusters.
%
%
 These properties together with high resolution imaging allowed observers to find and study them in many types of galaxies: interacting/merging systems (e.g. {\cite{Whitmore1}} and aseveral subsequent papers on the Antennae galaxies;  {\cite{deGrijs03}};  {\cite{Bastian05}};  {\cite{Trancho07}}; etc.) and dwarf starburst galaxies (e.g.  {\cite{ostlin03}};  {\cite{Anders04}}), LIRGs ({\cite{Wilson06}};  {\cite{Vaisaien}}) and also spirals ( e.g. {\cite{Larsen02}}).  This implies that the SCs formation is a natural product of the star formation process and it is very efficient in merging/interacting starburst systems.  

\begin{figure}[h]
\begin{center}
\includegraphics[width=12cm]{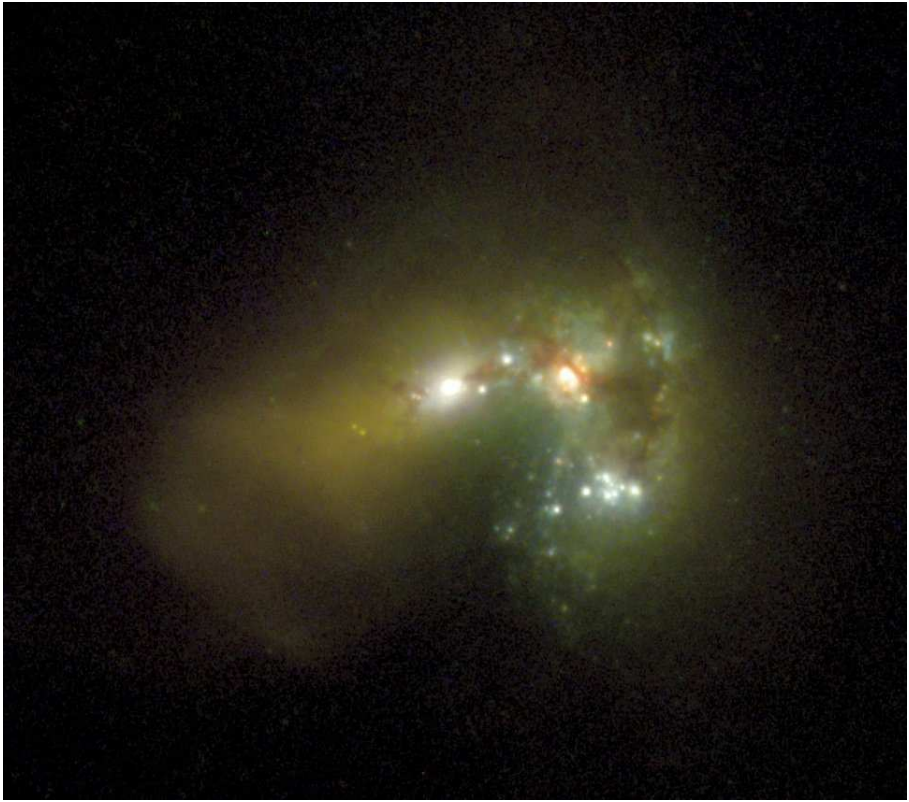}
\caption{Three-color image of Haro11 (ESO 350-IG 038). In red WFPC2/f814w filter, in green ACS/f435w filter and in blue ACS/f220w filter. The galaxy shows three starburst regions with numerous clusters and unrelaxed kinematic probably due to a past merger ({\cite{ostlin01}}). The central knot appears obscured by dust. It is a known Ly$\alpha$ and Ly$\alpha$-continuum emitter ({\cite{hayes07}} and thereafter.) }
\label{case2_1_ent}
\end{center}
\end{figure} 

Most of the star formation in a galactic system occurs in clustered environments through the collapse of giant molecular clouds. However, a large fraction of those systems are already unbound when they emerge from their dust cocoons ({\cite{Lada2}}).  In order to define the impact of bound cluster formation in the total star formation process, a parameter has been defined: the cluster formation efficiency, $\Gamma$. $\Gamma$ is the ratio between the cluster formation rate (CFR) and the total star formation rate (SFR) in the galaxy. Lada \& Lada ({\cite{Lada2}}) found a value of $\Gamma = 0.04 - 0.07$ for the solar neighborhood and Gieles \& Bastian ({\cite{Gieles1}}) got $\Gamma = 0.02 - 0.04$ for the SMC. Recently, Bastian ({\cite{Bastian08}}), using a set of simulations, concludes that the observed and optically detected clusters represent the 8$\%$ of the total star formation of a galaxy and this factor is independent of the SFR in the galaxy. If this is true, it implies that the cluster formation history (CFH) in the galaxy is directly connected with its total star formation history (SFH). In fact, a narrow relation exists between the star formation rate (SFR) of the galaxy and the cluster population formed at each epoch ({\cite{Larsen02}},{\cite{Bastian08}}) showing the major impact of SCs into the host system.  We can conclude that SCs are reliable tracers of the evolution history and of the present-day properties of their parent galaxy. 

Moreover SCs are relatively easy objects to study because they can be modeled as single stellar populations (SSPs) evolving in time. 
\begin{figure}[h]
\begin{center}
\includegraphics[width=12cm]{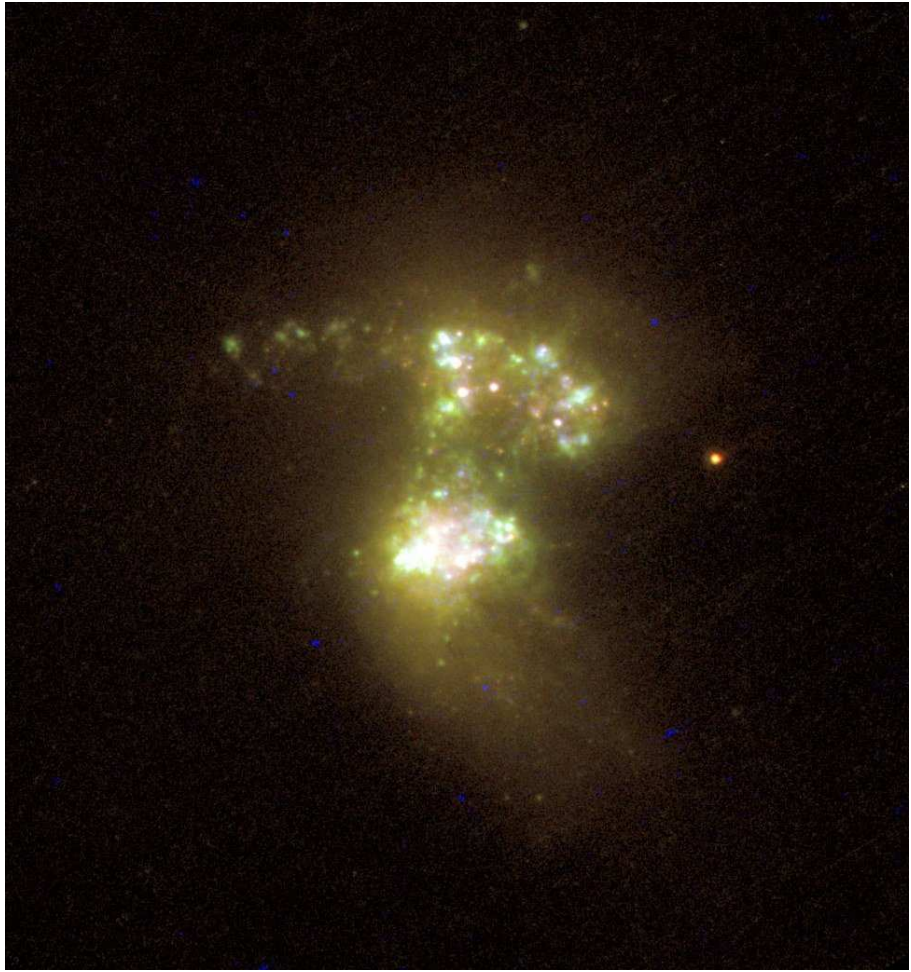}
\caption{Three-color image of MRK 0930 respectively. In red WFPC2/f814w filter, in green WFPC2/f606w filter, in blue WFPC2/f336w filter. Also in this case the perturbed morphology suggests that a merger event has been occurred. High cluster formation rate has been pointed out by \"Ostiln ( {\cite{ostlin00}}).}
\label{case2_1_ent}
\end{center}
\end{figure} 
This implies that starting with few initial conditions such as the IMF, the metallicity and a set of stellar evolutionary tracks evolving in time, we can use a SED (spectral energy distribution) fitting model to estimate ages, internal extinction, and masses of the clusters. 
It is clear that the presence of clusters in the BCGs represent a great opportunity for studying these galactic systems in detail. In 2007 we got a complete set of HST images ranging from the UV to IR for three different BCGs: ESO185, Haro11,  and MRK930. The amazing images (Figure 1-3) show that the systems have peculiar morphologies and hundreds of SCs which are the product of past, recent, and/or active starburst phases. What triggered the starbursts in these galaxies? When did this start?  Have the starbursts been continuously producing SCs or can we trace a few burst phases? Which rate have the SCs been produced? In these smaller galaxies, what is the fraction of the total mass that goes into clusters? Do BCGs contain also an old population of old star cluster? These are some of the questions we want to answer.

\section{Observational sample \& Data reduction}

The three targets have been observed with the Hubble Space Telescope (HST). A description of the available bandpass data are given in the tables 1 and 2. The data are mainly coming from the  observational program \#GO10902 and, in the case of Haro11, partially from other two programs \#GO9470 and \#GO10575. All the images have been reduced with the {\it IRAF/STSDAS}  task {\it MultiDrizzle} and aperture photometry performed with {\it DAOFIND/phot} task. 
\begin{table}
\caption{\label{arttype} Set of observations of the three targets.}
\footnotesize\rm
\begin{tabular*}{\textwidth}{@{}l*{15}{@{\extracolsep{0pt plus12pt}}l}}
\br
bandpass&&Instrument&ExpTime(ESO185, MRK930)&ExpTime(Haro11)\\
\mr
 F140LP&FUV-band&ACS/SBC&2600s&2700s\\
 F220W&NUV-band&ACS/HRC&&1513s\\
 F330W&U-band&ACS/HRC&&800s\\
 F336W&U-band&WFPC2&1200s&\\
 F435W&B-band&ACS/WFC&&680s\\ 
 F439W&B-band&WFPC2&800s&\\
 F550M&V'-band (medium)&ACS/WFC&&471s\\ 
 F606W&V-band (broad)&WFPC2&4000s&4000s\\
 F814W&I-band&WFPC2&4500s&5000s\\
 F160W&H-band&NIC3&5000s&5000s\\
\br
\end{tabular*}
\end{table}
The deep images in V (WFPC2/f606w, it's a broad V band) and I (WFPC2/f814w) have been used to capture most of the SC candidates in the three galaxies. We run {\it SExtarctor} separately in both the images and cross-correlatated  the two catalogues in order to remove all the spurious detections. After this selection process the catalogue of ESO185 contained more then 1000 clusters candidates, the catalogue of Haro11 and MRK930 more than 300 confirmed detections each. These results confirmed what we expected already looking at the impressive images of those systems: the cluster formation has been really efficient in those small starburst galaxies.

This catalogue of positions has been used for performing aperture photometry in all the available passbands. We corrected the magnitudes of the objects for the used aperture, galactic extinction, and the CTE (charge transfer efficiency) effect. We constrained completeness limit detection for both V and I images using the available tool BAOlab ({\cite{larsen99}}) to simulate sources to past into the original frames. We found 90 \% compliteness at $\sim$27.0 mag in both V and I images for all the three galaxies. This result looks promising because one of our goal is to detect and constraint the old star cluster populations.

To fix the final photometric catalogue of each galaxy, we removed all the sources with a photometric error in V and I filter larger than 0.2. This conservative cut has been done in order to better constraint the photometric properties of the clusters as we show in the next section.  

\section{Preliminary results}

What we are going to show in this section are the first results of our analysis. We would like to give some qualitative considerations on the SC population properties postponing a deep analysis in a coming paper.

In Figure 4 we show the color-magnitude and color-color plots of Haro11. Already from this simple plots we can see the presence of a large range of object ages, the youngest sitting in the bottom left part of the diagram, with bright M$_V$ magnitude and negative (V-I) color, and the older moving up to the right side, where M$_V$ are fainter and (V-I) color redder. In order to estimate ages and masses of the clusters we have used a SED (spectral energy distribution) fitting method. A description of the SED fitting code is found in Hayes et al. ({\cite{hayes07}}) based on {\it starburst99} spectral evolutionary models (Geneva's tracks). The input parameters are: metallicity $Z = 0.004$, Salpeter's IMF, with $\alpha=2.35$, in the mass range from 0.1 up to 120 $\msun $, instantaneous burst with a mass normalization of $ 10^{6} \msun $. From the fitting outputs it appears that the formation of SCs increased remarkably in the last 40 Myr in ESO185 (Figure 5). Moreover the distribution of the SC masses peaks at few $ 10^{4} \msun  $ with several clusters more massive than $ 10^{5} \msun $ (the range of the super star clusters). We conclude that  although BCGs are smaller than the massive merging, systems this doesn't prevent them from forming very massive star clusters. The next step will be to investigate the impact of these massive clusters on the star formation activity and the evolution of their host galaxy ({\cite{tenorio07}}, {\cite{Bastian08}}).

\begin{figure}[h]
\includegraphics[width=7.5cm]{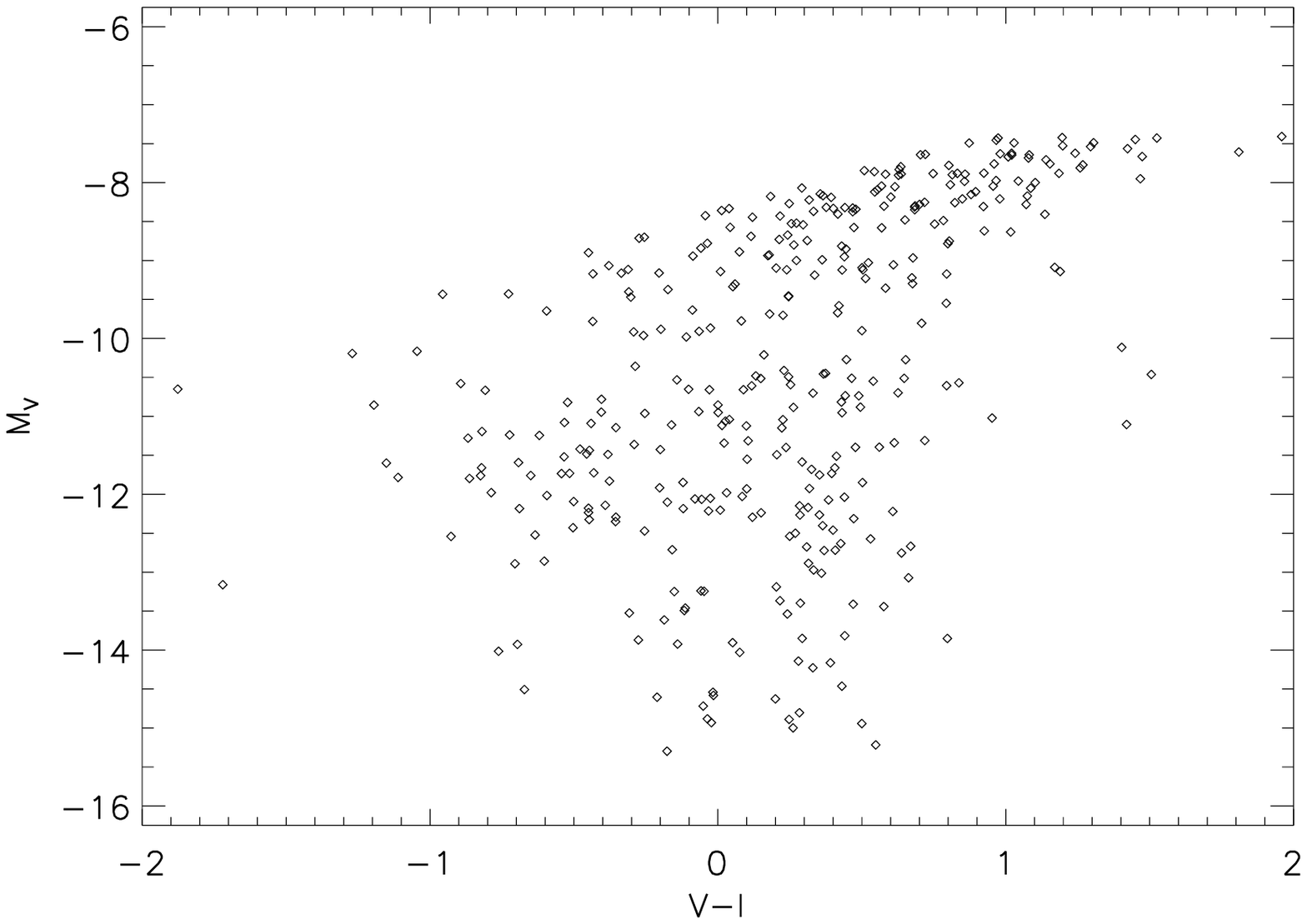}
\includegraphics[width=7.5cm]{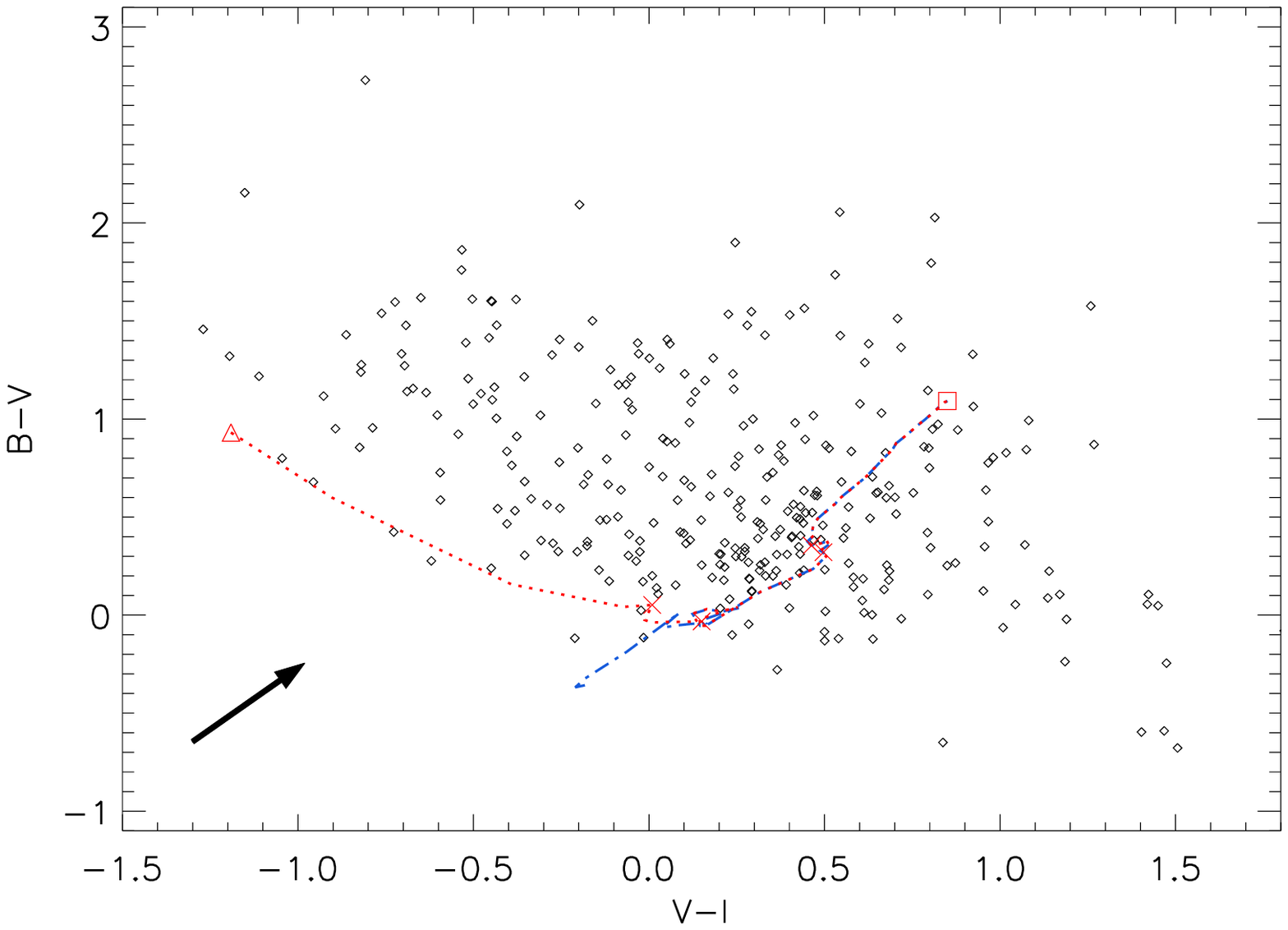}
\caption{Color plots for Haro11. On the left, M$_V$ versus (V-I) color. 90$\%$ completeness limit is 27.0 for both V and I. On the right, the (V-I) versus (B-V) color of the detected SCs. The dotted red line show the models by Zackrisson et al. {\cite{erik01}}.This model takes into account for stellar emissions and nebular emission, the latest crucial at the youngest phases. The triangle show the beginning of the model (age of $5\times10^5$ yrs) and the square the end at 14 Gyr. The stars along the model are located at  $5\times10^6$, $1\times10^7$, $5\times10^7$ and $1.5\times10^8$. If only stellar emission are taking in account (blue dot-dashed line, Marigo et al., {\cite{marigo08}}) we wouldn't be able to explain the SC colors in the left part of the plot. The arrow shows in which direction an extinction E(B-V)=0.32 moves the objects in the color-color plot. } 
\label{case2_1_ent}
\end{figure} 

However we can't neglect that the broad V filter suffers of contamination from nebular emission, which are really strong in the first 10 Myr in the life of a cluster. For this reason we decided to also use the spectral evolutionary model by Zackrisson et al. ({\cite{erik01}}) that takes into account both stellar spectra and gas continuum and line emissions.  In the Figure 4, the color-color plot on the right side shows the position of the Haro11 clusters and two different models, Zackrisson et al. ({\cite{erik01}}) with nebular contributions (red dotted line) and Marigo et al. ({\cite{marigo08}}) based on the Padova's tracks without considering the young nebular phase emission (blue dotted line). If only stellar emission is taking into account we wouldn't be able to explain the SC colors in the left part of the plot. Again from this color-color diagram we can distinguish between the young and the old SC populations and an internal dust reddening gradient.

\begin{figure}[h]
\includegraphics[width=7.5cm]{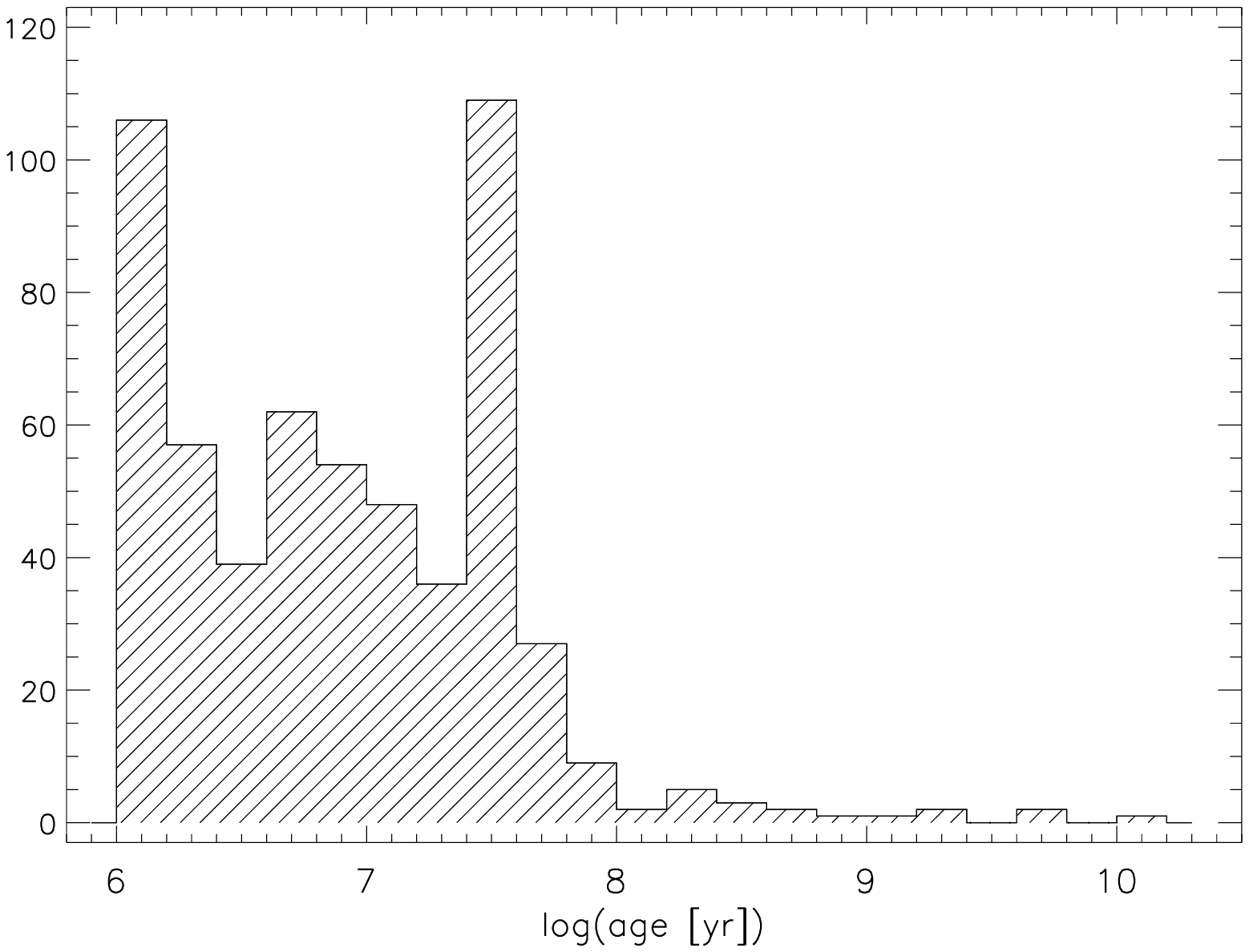}
\includegraphics[width=7.5cm]{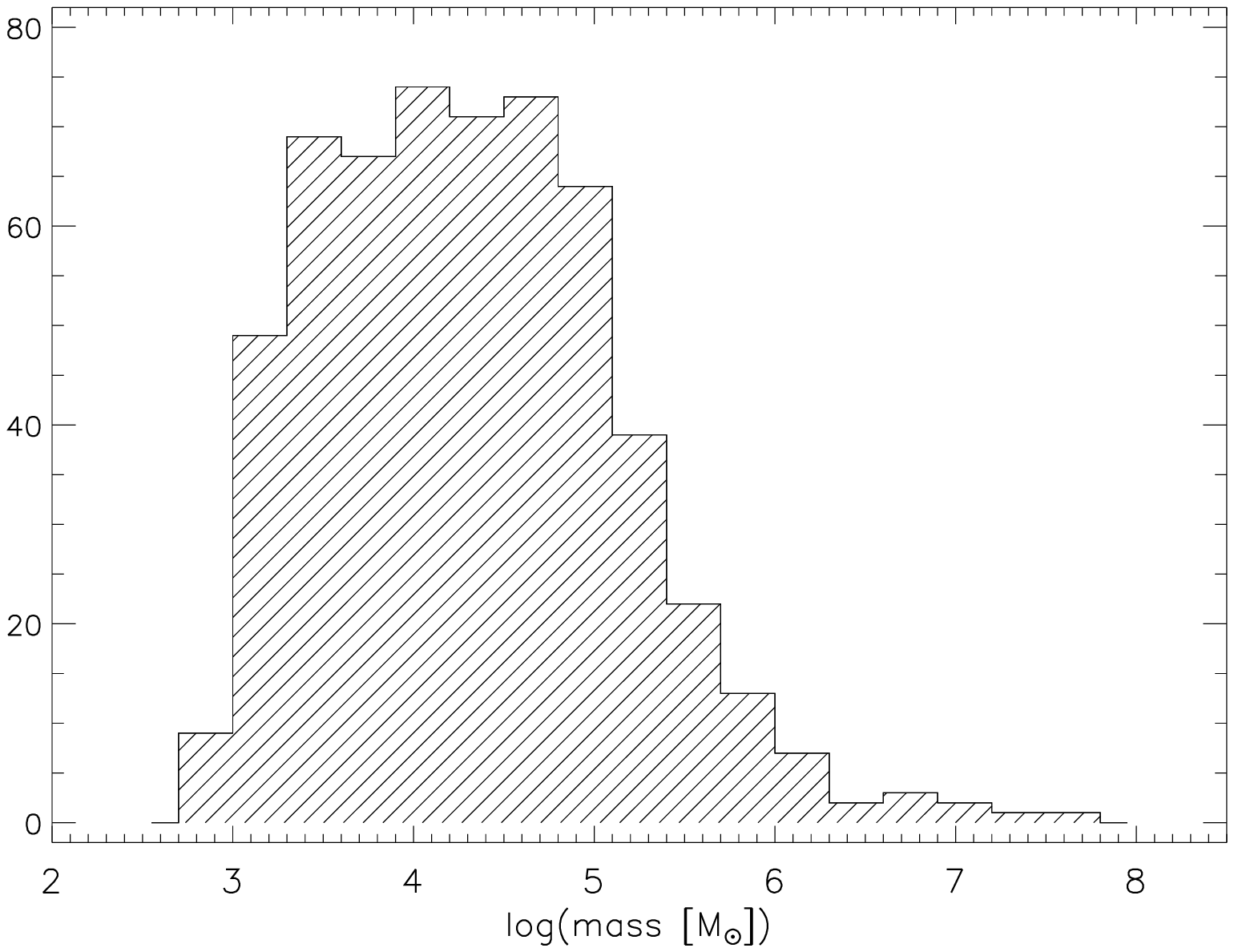}
\caption{Age (left) and mass (right) distributions of Eso185 SCs.  }
\label{case2_1_ent}
\end{figure} 

\section{Conclusions \& Open questions}
Our analysis is focused on the SC population of three BCGs. As pointed out in the introduction section, we want use the SC analysis in order to  study their host galaxy evolution and, more in general, the star formation mechanisms. In fact, SSCs have a major impact in the subsequent evolution of the galaxy. Moreover, BCGs represent peculiar objects in the wide range of galaxy types.  They are small systems with low metallicity content and high star formation rate so that to be classified as starburst galaxies. However, they contain also a red and old stellar population indicatives of a probable merger between an evolved dwarf galaxy with another gas-rich dwarf galaxy. 

Our three BCGs appear particularly rich in SCs. In the case of Haro11 and Eso185, their age and mass ranges are quite wide. Eso185 experienced an increment of the cluster formation rate in the last 40 Myr and several of those clusters are really massive. 

We will publish soon a complete statistical analysis on the SCs and on their corresponding galactic systems. We want to shed light on the spatial and temporal evolution of the starburst and also on the efficiency at which these BCGs produce bound clusters.  In the cases where it will be possible, we will also put constraints on the compactness of the SCs, estimating their effective radii.

\ack
A. Adamo thanks Robert Cumming and Genoveva Micheva for the useful discussions and for providing good corrections to this work. Special thanks are also for Angel R. Lopez-Sanchez who greatly improved the 3-colors images of the targets. We are really grateful to the organizers for all the efforts they have put into planning such an interesting conference.

\end{document}